\documentclass[aps,pre,twocolumn,groupedaddress,showpacs,floatfix,superscriptaddress]{revtex4-1}

\usepackage[T1]{fontenc}
\usepackage[utf8]{inputenc}
\usepackage{textcomp} 
\usepackage{graphicx}
\usepackage{amsmath}
\usepackage{amsfonts}
\usepackage{amssymb}
\usepackage{epsfig}
\usepackage{color}
\usepackage{bbm}
\usepackage{hyperref}
\usepackage{times}
\hypersetup{
    colorlinks = false,
    hidelinks=true
}

\usepackage{color}

\newcommand{\ep}{\mathcal{E}_p}
\newcommand{\ek}{\mathcal{E}_k}

\newcommand{\xidn}{\xi_{DN}}

\usepackage[colorinlistoftodos,prependcaption,textsize=small]{todonotes}

\begin{document}

\title{Energetics of the single-well undamped stochastic oscillators}

\author{Michał Mandrysz}
\email{michal.mandrysz@student.uj.edu.pl} \affiliation{Marian Smoluchowski
Institute of Physics, and Mark Kac Center for Complex Systems
Research, Jagiellonian University, ul. St. Łojasiewicza 11,
30--348 Kraków, Poland}

\author{Bartłomiej Dybiec}
\email{bartek@th.if.uj.edu.pl} \affiliation{Marian Smoluchowski
Institute of Physics, and Mark Kac Center for Complex Systems
Research, Jagiellonian University, ul. St. Łojasiewicza 11,
30--348 Kraków, Poland}

\date{\today}

\begin{abstract}
The paper discusses analytical and numerical results for non-harmonic, undamped, single-well, stochastic oscillators driven by additive noises.
It focuses on average kinetic, potential and total energies together with the corresponding distributions
under random drivings, involving Gaussian white, Ornstein-Uhlenbeck and Markovian dichotomous noises.
It demonstrates that insensitivity of the average total energy to the single-well potential type, $V(x) \propto x^{2n}$, under Gaussian white noise does not extend to other noise types.
Nevertheless, in the long-time limit ($t \to \infty$), the average energies grow as power-law with exponents dependent on the steepness of the potential $n$.
Another special limit corresponds to $n\to\infty$, i.e. to the infinite rectangular potential well, when the average total energy grows as a power-law with the same exponent for all considered noise types.
\end{abstract}


\maketitle


\section{Introduction\label{sec:introduction}}

Energy conservation is one of the fundamental principles of physics.
In deterministic models devoid of energy dissipation and external driving the energy conservation is typically built-in, whereas stochastic models usually require additional constraints to uphold it.
Stochastic systems in which such a behavior is desirable and successfully implemented by careful balancing of the dissipating and noise related terms are said to fulfill the fluctuation-dissipation theorem \cite{sekimoto2010stochastic,seifert2012stochastic}.
By definition, stochastic models conserve energy in a statistical sense and ought to represent physical states at the equilibrium or approaching it.
Regardless of their enormous success, the class of equilibrium and near-equilibrium systems hardly exhausts all observed possibilities.
In consequence, many models of non-equilibrium phenomena \cite{gammaitoni1998,doering1992,reimann2002, seifert2012stochastic, joubaud2007fluctuation} have been developed with a multitude of different approaches and methods \cite{hwalisz1989colored,metzler2000,metzler2004}.
In particular, considerable attention has been given to the task of classification
of various systems \cite{czopnik2003frictionless, bena2006, eab2018ornstein} described by stochastic differential equations in-and-out of equilibrium.

In the case of stochastic systems relevant to this work, a field worth of investigation is the regime of low or vanishing dissipation in the second-order Langevin (stochastic Newton) equation.
Accordingly, the lack of dissipation results in the growth of average energies due to the stochastic force.
As will be shown, the short time behavior of average energies is often very different from the long-time (asymptotic) characteristics.
Having withheld the fluctuation-dissipation theorem, one could enquire about the survival of other notions of statistical mechanics, such as the equipartition theorem \cite{reichl1998,reif2009}.
The paper shows that, among models studied here, the only case where despite of absence of stationary states, the equipartition of energy is asymptotically satisfied is the (undamped) harmonic stochastic oscillator which has been studied previously in many contexts \cite{lin2011undamped,mandrysz2018energetics}.
Nonetheless, we would like here to extend these earlier results and probe the properties of non-harmonic, undamped, stochastic oscillators and non-Gaussian noises \cite{chechkin2008introduction}.
In particular, a natural and simple extension to non-harmonic potentials includes symmetric single-well potentials of the following form:
\begin{equation}
	V(x)=k\frac{x^{2n}}{2n}.
	\label{eq:potential}
\end{equation}
Having narrowed the scope of our research, we start by shortly presenting the considered model in the damped (dissipating) context, indicating the way of obtaining
the quantities of interest and finally disregarding the damping term (Sec.~\ref{sec:model}).
For the sake of completeness, the case of a free particle (Sec.~\ref{sec:model-free}) is considered first.
Next, we proceed to restate the results mentioned above (Sec. \ref{sec:harmonic}) and extend this approach to calculate explicitly the
uncertainties connected with the mean values of energy.
From there we continue to the last part of this paper regarding the non-harmonic single-well potentials (Sec. \ref{sec:nonharmonic}).
The main characteristics of interest will be the asymptotic time dependence of energies (total, kinetic and potential).
The results are confirmed with numerical simulations for $n=1,2,3,\infty$ with the special attention to the case of $n=\infty$ and to the Markovian dichotomous noise \cite{bena2006} not considered explicitly in \cite{mallick2005anharmonic}.
Surprisingly, the solutions hold for colored noises different in nature  such as Ornstein-Uhlenbeck, dichotomous noise, and can give identical long-time behavior if only their correlation times match.
The paper is closed with Summary and Discussion (Sec.~\ref{sec:summary}).

\section{Model and Results\label{sec:model}}

In what follows, we study properties of general stochastic oscillator, i.e. the motion in the single-well potential given by Eq.~\eqref{eq:potential} with $k>0$ and $n\in\{1,2,\dots\}$.
Nevertheless, in more general situations, it is also possible to consider non-integer $n>0$, in such a case it is necessary to replace $x$ with $|x|$.
The special case of $n=1$ corresponds to the harmonic oscillator which is one of fundamental models in statistical physics \cite{gitterman2005noisy,gitterman2013noisy}. Following the convention of \cite{kubo1966fluctuation,risken1984}, in the dimensional units, the evolution of the state variable $x(t)$ is described by the second order Langevin equation
\begin{equation}
m\frac{d^2x(t)}{dt^2}=-\gamma m  \frac{dx(t)}{dt} - kx^{2n -1}(t) + \sqrt{2 \gamma k_B T m}\xi(t),
\label{eq:full-langevin}
\end{equation}
where $x(t)$ represents the position, $m$ the particle mass, $T$ the system temperature, $k_B$ Boltzmann constant and $\gamma$ is a damping coefficient.
In Eq.~(\ref{eq:full-langevin}) $\xi(t)$ stands for the Gaussian white noise (GWN) satisfying
\begin{equation}
\langle \xi(t) \rangle=0
\label{eq:gwn-av}
\end{equation}
and
\begin{equation}
\langle\xi(t) \xi(s) \rangle= \delta(t-s).
\label{eq:gwn-cor}
\end{equation}

In addition to the Gaussian white noise (GWN), we also consider the symmetric Markovian dichotomous noise (DN) and Ornstein-Uhlenbeck noise (OUN).
The symmetric, allowing two possible values $\pm 1$, Markovian dichotomous noise $\xi_{DN}(t)$, see \cite{horsthemke1984,bena2006}, satisfies
\begin{equation}
 \langle \xi_{DN}(t) \rangle = 0
 \label{eq:dn-average}
\end{equation}
and
\begin{equation}
 \langle \xi_{DN}(t)\xi_{DN}(s) \rangle =  \exp\left[-2\lambda|t-s|\right],
 \label{eq:dn-ac}
\end{equation}
where $\lambda$ is the transition rate between states \cite{horsthemke1984}.
The Ornstein-Uhlenbeck noise (OUN) \footnote{This definition does converge to the GWN for $\mathcal{D}=\rho$ and $\rho \to \infty$.} is the process defined by the following Langevin equation
\begin{equation}
 \frac{d\xi_{OU}(t)}{dt}=-\rho \xi_{OU}(t) + \mathcal{D} \xi(t),
 \label{eq:ou-langevin}
\end{equation}
where $\xi(t)$ is the Gaussian white noise, see Eqs.~(\ref{eq:gwn-av}) -- (\ref{eq:gwn-cor}).
However, in numerical simulations it is often more convenient to use the so-called exact updating formula for the OUN \cite{gillespie1996exact}, rather than solving the Eq.~\eqref{eq:ou-langevin} directly.
Moreover, the OU process fulfills
\begin{equation}
 \langle \xi_{OU}(t) \rangle = 0
 \label{eq:ou-average}
\end{equation}
and
\begin{equation}
 \langle \xi_{OU}(t)\xi_{OU}(s) \rangle = \frac{\mathcal{D}^2}{2\rho}\exp\left[-\rho|t-s|\right],
 \label{eq:ou-ac}
\end{equation}
under condition that $\xi_{OU}(-\infty)=0$.

The Gaussian white noise describes interactions of the oscillator with the thermal bath of temperature $T$.
Langevin equation~(\ref{eq:full-langevin}) is the Newton second law accounting for a random force $\xi(t)$.
It describes the system evolution on the microscopic level.
Both variables position $x(t)$ and velocity  $v(t)=\dot{x}(t)$ are no longer deterministic, but become random variables distributed according to some probability density $P(x,v;t)$.
The probability of finding the system in a state characterized by $(x(t),v(t))$ evolves according to the diffusion (Fokker-Planck) equation \cite{kubo1966fluctuation,risken1984}
\small
\begin{equation}
 \partial_t P(x,v;t) =\left[\partial_v\left( \gamma v + \frac{V'(x)}{m}   \right) - v \partial_x  +\gamma \frac{k_B T}{m}\partial^2_v \right]P(x,v;t).
 \label{eq:kk}
\end{equation}
\normalsize
For any potential $V(x)$, such that $V(x) \to \infty$ as $x\to\pm\infty$, the stationary solution of Eq.~(\ref{eq:kk}) is of the Boltzmann-Gibbs type
\begin{equation}
 P(x,v) \propto \exp\left[ - \frac{1}{k_B T} \left(  \frac{m v^2}{2} + {V(x)} \right) \right].
 \label{eq:st}
\end{equation}
The exponent in Eq.~(\ref{eq:st}) is the total energy $\mathcal{E}$ which is the sum of kinetic $\mathcal{E}_k$ and potential $\mathcal{E}_p$ energies.
The system's total energy $\mathcal{E}=\mathcal{E}_k+\mathcal{E}_p=\frac{1}{2}mv^2+k\frac{x^{2n}}{2n}$ depends on its state $(x(t),v(t))$. Consequently, instantaneous energies, analogous to state variables, are random variables.
Nevertheless, average energies are constant for large $t$ due to the existence of a stationary state.
In the stationary state, the position and the velocity are statistically independent.
Finally, Eq.~(\ref{eq:full-langevin}) assures, that the stochastic harmonic oscillator, corresponding to $n=1$, fulfills the equipartition theorem \cite{kubo1966fluctuation,risken1984}.

For the purpose of deriving the quantities of interest, Eq.~(\ref{eq:full-langevin}) can be rewritten as a set of two first-order equations
\begin{equation}
\left\{
\begin{array}{ccl}
\frac{dx(t)}{dt} & = & v(t) \\
\frac{dv(t)}{dt} & = & -\gamma   v(t) - \omega^2 x^{2n -1}(t) + \sqrt{\frac{2 \gamma k_B T}{m}}\xi(t)
\end{array}
\right.,
\label{eq:set}
\end{equation}
where $\omega^2=k/m$.
For the parabolic potential these equations are linear, thus standard methods of solving linear differential equations can be applied \cite{risken1984,mao2007stochastic}. The system described by Eq.~(\ref{eq:full-langevin}) or Eq.~(\ref{eq:set}) in the presence of simple noises can be studied analytically \cite{risken1984,mao2007stochastic,tome2015stochastic,czopnik2003frictionless}.
From Eq.~(\ref{eq:set}) one can derive equations for moments $\langle v^2(t) \rangle$ and $\langle x^2(t) \rangle$ from which the evolution of average energies can be calculated.
The time evolution of average potential and kinetic energies are described by
\begin{equation}
 \frac{d}{dt} \langle \mathcal{E}_p(t) \rangle = k \langle v(t) x^{2n-1}(t) \rangle
 \label{eq:ep}
\end{equation}
and
\small
\begin{equation}
 \frac{d}{dt} \langle \mathcal{E}_k(t) \rangle = -\gamma m \langle {v^2(t)} \rangle -  k \langle v(t) x^{2n-1}(t) \rangle + \sqrt{2 \gamma k_B T m} \langle \xi(t) v(t) \rangle.
 \label{eq:ek}
\end{equation}
\normalsize
Finally, the total mechanical energy $\langle \mathcal{E} (t) \rangle$ varies in time according to
\begin{equation}
 \frac{d}{dt} \langle \mathcal{E}(t) \rangle = -\gamma m \langle {v^2(t)} \rangle  + \sqrt{2 \gamma k_B T m} \langle \xi(t) v(t) \rangle.
 \label{eq:e}
\end{equation}

Energy distribution $f(\ep,\ek)$ can be calculated by the transformation of variables
\begin{eqnarray}
 f(\ep,\ek) & = & f(x(\ep),v(\ek)) \times |\mathbf{J}| \nonumber \\
 & = &  \sum_{\{\pm\}} f\left(\pm \sqrt[2n]{\frac{2n\ep}{k}}, \pm \sqrt{\frac{2\ek}{m}} \right) \times |\mathbf{J}|,
 \label{eq:transformationofvariables}
\end{eqnarray}
where
\begin{equation}
 \mathbf{J} = \frac{\partial(x(\ep),v(\ek))}{\partial(\ep,\ek)}
\end{equation}
is the Jacobian of the transformation from $(x(t),v(t)) \to (\ep(t),\ek(t))$.
The sum in Eq.~(\ref{eq:transformationofvariables}) indicates summing over all combination of signs.
Two dimensional density  $f(\ep,\ek)$ is defined for $\ep \geqslant 0$ and $\ek \geqslant 0$ while $f(x,v)$ is defined on the whole plane.
From Eq.~(\ref{eq:transformationofvariables}) further properties of the energy distributions can be determined, including marginal densities and other characteristics.
Nevertheless, in the majority of situations, knowledge of the full $f(x,v)$ density is required.

In the absence of dissipation, i.e. when the $-\gamma v(t)$ term is disregarded, Eqs.~(\ref{eq:ep}),~(\ref{eq:ek}) and ~(\ref{eq:e}) further simplify.
Moreover, for special noise types exact solutions of these equations can be provided.
In such a case Eq.~(\ref{eq:set}) transforms into
\begin{equation}
\frac{d^2x(t)}{dt^2}= - \omega^2 x^{2n-1}(t) + \sqrt{h}\xi(t),
\label{eq:langevin}
\end{equation}
where $h={2 \gamma k_B T}/{m}$ is the independent parameter scaling the noise strength.
In further sections, we examine the undamped motion only, i.e. the system described by Eq.~(\ref{eq:langevin}).

%
%
\subsection{Free particle \label{sec:model-free}}

By direct calculation, it is possible to show that the average kinetic energy $\langle \ek(t) \rangle$, which is equal to the average full energy $\langle \mathcal{E}(t) \rangle$, scales linearly in time. The velocity can be calculated as
\begin{equation}
 v(t)=\int_0^t \xi(u) du.
\end{equation}
In particular, for a free particle driven by the Gaussian white noise one has:
\begin{eqnarray}
 \langle v(t) v(s) \rangle &  = &  \Big\langle \int_0^t \xi(u) du \int_0^s \xi(v) dv \Big\rangle \nonumber \\
 &  = &   \int_0^t du\int_0^s dv  \langle \xi(u)   \xi(v) \rangle \nonumber \\
 &  = &   \int_0^t du\int_0^s dv  \delta(u-v).   \nonumber \\
\end{eqnarray}
Finally, for $v(0)=0$, one obtains
\begin{equation}
 \langle v^2(t) \rangle = t.
\end{equation}
The average kinetic energy grows like
\begin{equation}
 \langle \ek (t) \rangle = \langle \mathcal{E}(t) \rangle = \frac{m}{2} \times t.
 \label{eq:gwn-free}
\end{equation}
As will be shown in the forthcoming subsections, the (long-time) evolution of the average total energy is the same for any single-well potential under the GWN, see Eq.~(\ref{eq:etot}) and left panel of Fig.~\ref{fig:n2wn-all} and Fig.~\ref{fig:GWNgeneral}.

For a free particle driven by the Markovian dichotomous noise one obtains
\begin{eqnarray}
 \langle v(t) v(s) \rangle &  = &  \Big\langle \int_0^t \xidn(u) du \int_0^s \xidn(v) dv \Big\rangle \nonumber \\
 &  = &   \int_0^t du\int_0^s dv  \langle \xidn(u)   \xidn(v) \rangle \nonumber \\
 &  = &   \int_0^t du\int_0^s dv \exp\left[ -\lambda |u-v| \right].   \nonumber \\
\end{eqnarray}
and
\begin{equation}
 \langle v^2(t) \rangle = \frac{-1+\exp(-2t\lambda)+2t\lambda}{2\lambda^2}.
 \label{eq:v2DN}
\end{equation}
Thus, asymptotically, one gets
\begin{equation}
 \langle v^2(t) \rangle \propto \frac{1}{\lambda} \times t
\end{equation}
and
\begin{equation}
 \langle \ek(t) \rangle \propto \frac{m}{2\lambda} \times t.
 \label{eq:free-dn}
\end{equation}

Analogous  calculations can be performed for the Ornstein-Uhlenbeck noise (from now we set $\mathcal{D}=1$ unless otherwise stated) resulting in
\begin{equation}
 \langle v^2(t) \rangle = \frac{-1+\exp(-t\rho)+t\rho}{\rho^3}
\end{equation}
and asymptotic formulas
\begin{equation}
 \langle v^2(t) \rangle \propto \frac{1}{\rho^2} \times t,
\end{equation}
\begin{equation}
 \langle \ek(t) \rangle \propto \frac{m}{2\rho^2} \times t.
\end{equation}

For a free particle, $V(x)=0$, the total energy $\mathcal{E}$ is given by the kinetic energy $\ek$.
Energy distribution can be calculated by the change of variables
\begin{eqnarray}
f(\mathcal{E})& = &  f(\ek) \nonumber \\
& =& \int_{-\infty}^\infty f(x,v(\ek))  \left| \frac{d v}{ d \ek} \right| dx \nonumber \\
& = & \frac{\sqrt{2}}{m\sqrt{\ek}}  \int_{-\infty}^\infty f\left(x,\sqrt{\frac{2\ek}{m}}\right) dx \nonumber \\
& = & f(v(\ek))\frac{\sqrt{2}}{\sqrt{m\ek}}.
\end{eqnarray}
For Gaussian white and Ornstein-Uhlenbeck noises,
due to the Gaussian distribution of random pulses,
$f(x,v)$ distributions are two dimensional (2D) normal densities for which the correlation matrix can be calculated by standard methods \cite{risken1984}.
Nevertheless, for the free particle, the knowledge of the correlation matrix is not necessary to derive the energy distribution because it is enough to know the marginal density, which is Gaussian.
The kinetic energy distribution, as well as the full energy distribution, has the same functional dependence as the distribution of the kinetic energy for the harmonic ($n=1$) potential, see below.
The non-trivial parameter of the energy distribution is the average energy which, for a free particle, follows a different scaling than for the harmonic potential, e.g. compare Eq.~(\ref{eq:gwn-free}) and (\ref{eq:ek_wn}).

%
%
\subsection{Harmonic potential ($n=1$) \label{sec:harmonic}}

\subsubsection*{Average energies}

For the harmonic ($n=1$) potential with $x(0)=0$, $v(0)=0$  and the Gaussian white noise appropriate integrals can be performed, see \cite{mandrysz2018energetics}, resulting in
\begin{equation}
\langle \mathcal{E}_k(t) \rangle = h \frac{2\omega t + \sin (2 \omega t)}{8 \omega },
\label{eq:ek_wn}
\end{equation}
\begin{equation}
\langle \mathcal{E}_p(t) \rangle = h \frac{2\omega t - \sin (2 \omega t)}{8 \omega },
\label{eq:ep_wn}
\end{equation}
and
\begin{equation}
\langle \mathcal{E}(t) \rangle = \frac{h}{2} \times t,
\label{eq:e_wn}
\end{equation}
where $\omega=\sqrt{k/m}$.

Due to the lack of the damping term, $-\gamma v(t)$, the Gaussian white noise pumps energy into the system.
For a sufficiently large $t$ approximately half of the total energy is stored as kinetic one, while the remaining half is stored as the potential energy
\begin{equation}
 \langle \mathcal{E}_k (t) \rangle \simeq \frac{1}{2} \langle \mathcal{E} (t) \rangle
\end{equation}
and
\begin{equation}
\langle \mathcal{E}_p (t)\rangle \simeq \frac{1}{2} \langle \mathcal{E} (t) \rangle .
\end{equation}
With increasing $t$ the quality of this approximation
increases and the approximation becomes exact as $ t \rightarrow \infty $.
If the GWN is replaced by the symmetric Markovian dichotomous noise $\xi_{DN}(t)$,
see Eqs.~(\ref{eq:dn-average}) and~(\ref{eq:dn-ac}), one can also calculate average energies.
For $\xi_{DN}(0) \in \{- 1,+1\}$ with probability $1/2$, formulas for average energies
can be found in \cite{mandrysz2018energetics}.
%
The asymptotic (large $t$) formula for the average total energy takes the following form
\begin{equation}
\langle \mathcal{E}(t) \rangle \propto  \frac{2 \lambda  h}{4 \lambda ^2+\omega ^2} \times t.
\label{eq:e_dn_as}
\end{equation}
For the average kinetic energy we get
\begin{equation}
\langle \mathcal{E}_k(t) \rangle \propto \frac{\lambda  h}{4 \lambda ^2+\omega ^2} \times t \simeq \frac{1}{2}\langle \mathcal{E}(t) \rangle
\label{eq:ek_dn_as}
\end{equation}
and for the average potential energy
\begin{equation}
\langle \mathcal{E}_p(t) \rangle \propto \frac{\lambda  h}{4 \lambda ^2+\omega ^2} \times t \simeq \frac{1}{2}\langle \mathcal{E}(t) \rangle.
\label{eq:ep_dn_as}
\end{equation}
Average energies, analogously like for the GWN, grow linearly in time.

Exact formulas can be also derived for the OUN replacing the GWN, see \cite{mandrysz2018energetics}.
Asymptotically, average energies grow like
\begin{equation}
\langle \mathcal{E}_k(t) \rangle \propto \frac{h}{4\left(\rho ^2+\omega ^2\right)} \times t \simeq \frac{1}{2}\langle \mathcal{E}(t) \rangle,
\label{eq:ek_oun_as}
\end{equation}
\begin{equation}
\langle \mathcal{E}_p(t) \rangle \propto \frac{h}{4\left(\rho ^2+\omega ^2\right)} \times t \simeq \frac{1}{2}\langle \mathcal{E}(t) \rangle
\label{eq:ep_oun_as}
\end{equation}
and
\begin{equation}
\langle \mathcal{E}(t) \rangle \propto \frac{h}{2\left(\rho ^2+\omega ^2\right)} \times t .
\label{eq:e_oun_as}
\end{equation}

The long-time behavior of the stochastic harmonic oscillator  driven by simple noises,
e.g. Gaussian white noise (GWN), Markovian dichotomous noise (DN) or Ornstein-Uhlenbeck noise (OUN)
was studied in \cite{mandrysz2018energetics} where exact formulas are provided.
From Tab.~\ref{tab:sumuup} and Eqs.~(\ref{eq:ek_wn}) -- (\ref{eq:e_oun_as}) it is clearly visible that asymptotically average energies grow linearly in time.
Moreover, the average total energy is equally divided between average kinetic and potential energies, i.e. equipartition of energy is fulfilled.
\begin{table}[!htbp]
\begin{center}
\begin{tabular}{l||c|c}
	noise
	& $\underset{t \to \infty}{\lim} \langle \mathcal{E}(t) \rangle $
	& $\underset{t \to \infty}{\lim} \langle \ek(t) \rangle/\langle \ep(t) \rangle$\\
	\hline\hline
	GWN 			& $\frac{h}{2} \times t$									& 1\\
	DN				& $\frac{2 \lambda  h}{4 \lambda ^2+\omega ^2} \times t$	& 1\\
	OUN				& $\frac{h}{2\left(\rho ^2+\omega ^2\right)} \times t	$	& 1\\

\end{tabular}
\end{center}
\caption{Asymptotic dependence of the average energy $\langle \mathcal{E} (t) \rangle$  and the ratio of average energies  $\langle \ek(t) \rangle/\langle \ep(t) \rangle$
for the harmonic oscillator driven by various noise types.}
\label{tab:sumuup}
\end{table}
\subsubsection*{Energy distributions}

For the harmonic potential the deterministic force $-V'(x)=-k x$ is linear, consequently for the GWN driving the 2D probability density
$f(x,v)$ is a 2D, time dependent, normal density.
Similarly, time dependent marginal densities $f(x)$ and $f(v)$ are 1D Gaussians with parameters determined from the full $f(x,v)$ density, i.e. $\langle x(t) \rangle$, $\sigma^2(x(t))$ and $\langle v(t) \rangle$, $\sigma^2(v(t))$.

\begin{figure}[!h]%
\centering
\includegraphics[angle=0,width=0.9\columnwidth]{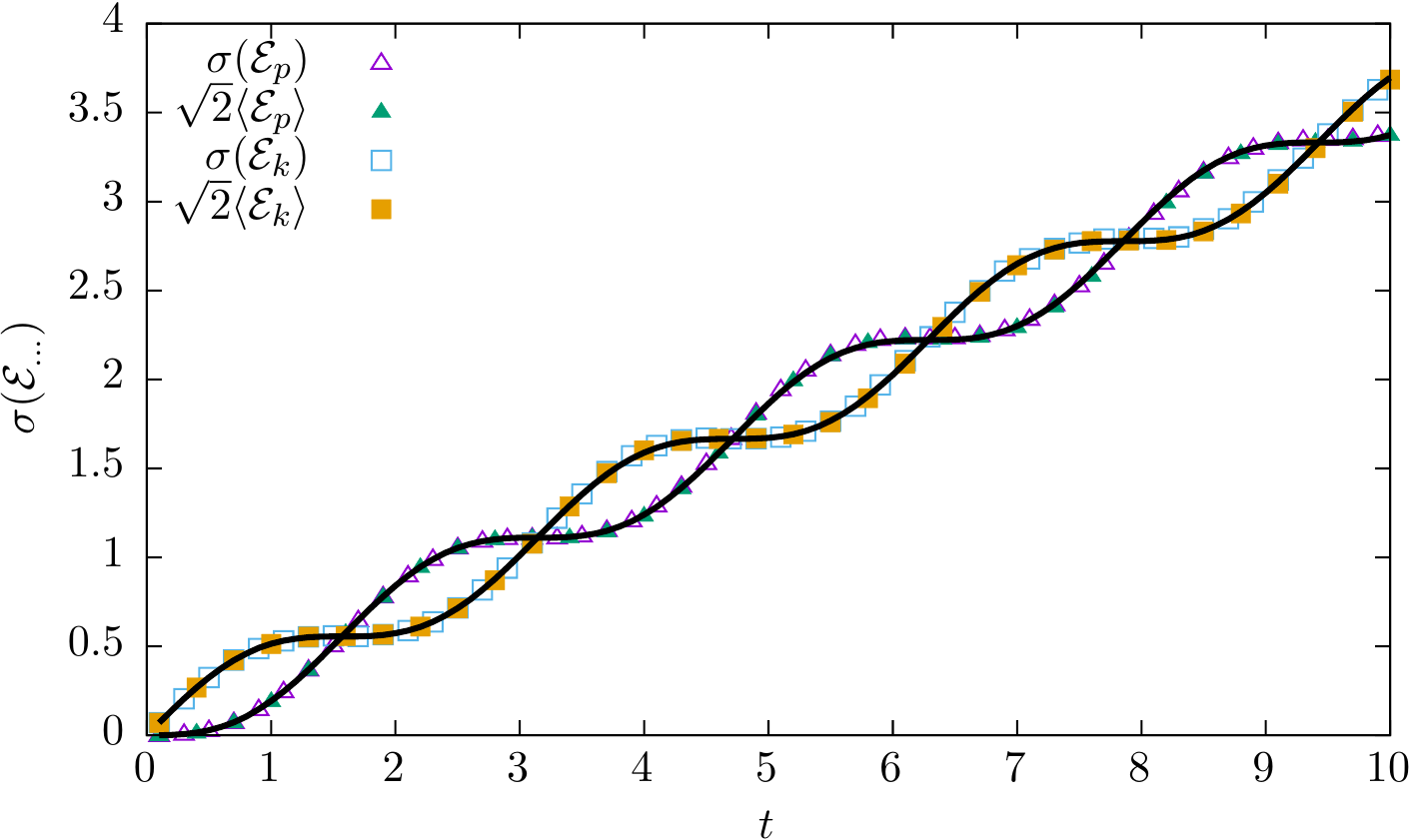} \\
\caption{Standard deviations of kinetic, $\sigma( \mathcal{E}_k(t) )$ and potential, $\sigma (\mathcal{E}_p(t) )$, energies as a function of time
for the harmonic potential perturbed by the Gaussian white noise.
Points correspond to results of simulations, while
solid lines present theoretical formulas (\ref{eq:ep-sigma}) and~(\ref{eq:ek-sigma}), i.e.
Eqs.~(\ref{eq:ek_wn}) -- (\ref{eq:ep_wn}) multiplied by $\sqrt{2}$.
More detail in the text.
}
\label{fig:n2wn-energy}
\end{figure}



\begin{widetext}

	\begin{figure}[!h]%
	\centering
\includegraphics[angle=0,width=1.0\columnwidth]{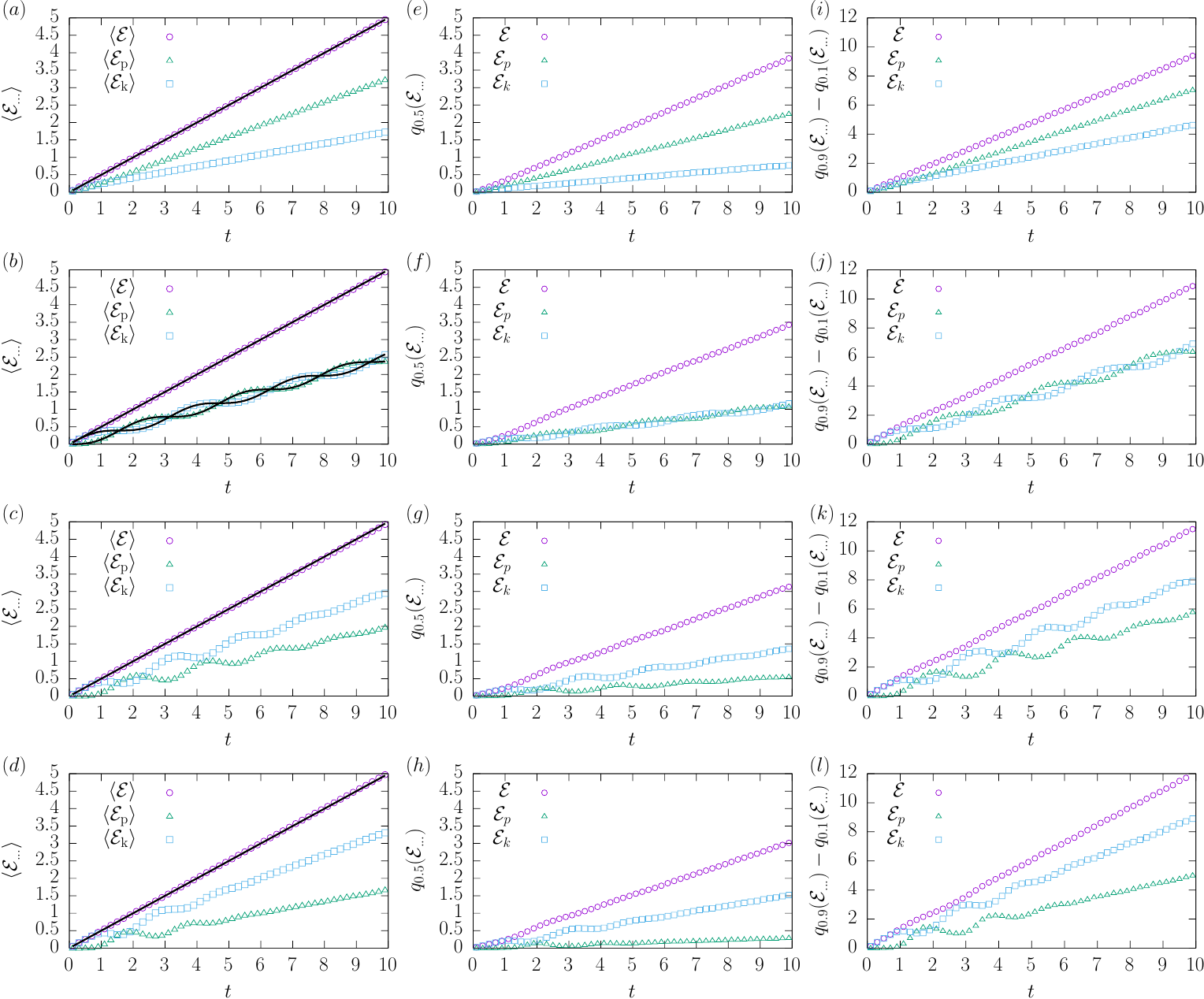} \\
	\caption{Average energies $\langle \mathcal{E}_{\dots} (t) \rangle$ (left column), medians of energy distributions $q_{0.5}(\mathcal{E}_{\dots}(t))$ (middle column), inter-quantile widths of energy distributions $q_{0.9}(\mathcal{E}_{\dots}(t)) -q_{0.1}(\mathcal{E}_{\dots}(t))$ (right column) for linear (top panel), parabolic, cubic and quartic (bottom panel) potentials subject to the action of the Gaussian white noise. Solid lines present exact formulas given by Eqs.~(\ref{eq:ek_wn}) -- (\ref{eq:e_wn}) and~(\ref{eq:etot}).
	}
	\label{fig:n2wn-all}
	\end{figure}

\end{widetext}


The time dependent energy distribution $f(\ep,\ek)$ can be calculated by the transformation of variables, see Eq.~(\ref{eq:transformationofvariables}), with the Jacobian
\begin{equation}
 |\mathbf{J}| =  \left| \frac{\partial(x(\ep),v(\ek))}{\partial(\ep,\ek)} \right| =\frac{1}{2\sqrt{m k} \sqrt{\ep \ek}}.
\end{equation}
Taking into account that $\pm x$ gives the same $\ep$ and $\pm v$ results in the same $\ek$, one gets
\begin{eqnarray}
 f(\ep,\ek) & = &  f(x(\ep),v(\ek))  \times |\mathbf{J}| \nonumber \\
 & = & \sum_{\{\pm\}} f\left(\pm\sqrt{\frac{2\ep}{k}},\pm\sqrt{\frac{2\ek}{m}}\right) \times | \mathbf{J}|.\end{eqnarray}
Marginal densities are defined in the standard manner
\begin{equation}
 f(\ep)=\int_0^\infty f(\ep,\ek) d\ek,
\end{equation}
\begin{equation}
 f(\ek)=\int_0^\infty f(\ep,\ek) d\ep.
\end{equation}

In a general situation, in order to find marginal densities, it is necessary to find the 2D $f(\ep,\ek)$ density first and then perform the appropriate integration.
Since $\ep$ depends on the position $x$ only and $\ek$ depends on the velocity $v$ only, the transformation does not mix variables and the marginal densities of $\ep$ and $\ek$ can be calculated from the marginal densities of $x$ and $v$.

For instance, at a given time point $t$
\begin{eqnarray}
 f(\ep)& = & \int_0^\infty f(\ep,\ek) d\ek \nonumber \\
 & = & \int_0^\infty f(x(\ep),v(\ek))  \frac{2}{\sqrt{m k} \sqrt{\ep \ek}} d\ek \nonumber \\
 & = & \int_0^\infty f(x(\ep),v)  \frac{2}{\sqrt{k \ep}} \frac{2 dv}{\sqrt{2}}\nonumber \\
& = & \int_{-\infty}^\infty f(x(\ep),v)  \frac{\sqrt{2}}{\sqrt{k \ep}} dv \nonumber \\
  & = & f(x(\ep)) \frac{\sqrt{2}}{\sqrt{k \ep}},
\end{eqnarray}
where $\frac{d\ek}{\sqrt{m\ek}}=\frac{2 dv}{\sqrt{2}}$.

Let us calculate $\langle \ep(t) \rangle $ and $\sigma^2(\ep(t))$.
The marginal density $f(x)$, as a marginal density of 2D normal distribution, is 1D Gaussian density
\begin{equation}
 f(x)=\frac{1}{\sqrt{2\pi \sigma^2(x(t))}} \exp\left[ -\frac{(x-\langle x (t) \rangle )^2}{2\sigma^2(x(t))} \right],
\end{equation}
where $\langle x (t) \rangle =0$ and
\begin{equation}
 \sigma^2(x(t))=\frac{2}{k}\langle \ep(t) \rangle.
\end{equation}
The marginal distribution of $\ep$ is
\begin{eqnarray}
 f(\ep) & = &  \frac{1}{\sqrt{4\pi \langle \ep (t) \rangle/k}} \exp\left[ -\frac{\ep }{2\langle \ep (t) \rangle} \right] \times \frac{\sqrt{2}}{\sqrt{k \ep}} \nonumber \\
 & = & \frac{1}{\sqrt{2\pi \langle \ep (t) \rangle}} \exp\left[ -\frac{\ep }{2\langle \ep (t) \rangle} \right] \times \frac{1}{\sqrt{\ep }}
\end{eqnarray}
and its cumulative density is
\begin{equation}
 \mathcal{F}(\ep)=\mathrm{erf}\left[  \sqrt{\frac{\ep}{2\langle \ep (t) \rangle}} \; \right],
 \label{eq:ep-ccdf}
\end{equation}
where $\mathrm{erf}()$ is the error function and $\langle \ep (t) \rangle$ is given by Eq.~(\ref{eq:ep_wn}).
Eq.~(\ref{eq:ep-ccdf}) gives the cumulative density (CDF) from which the complementary cumulative distribution function (CCDF),
$1-\mathcal{F}(\dots)$, is obtained.
The CCDF is calculated at a time $t$ at which the average potential energy is $\langle \mathcal{E}_p (t) \rangle$, see Eq.~(\ref{eq:ep_wn}).
Indeed, as expected and confirmed by the integration, the mean value of $\ep$ at a time $t$ is
\begin{equation}
 \int_0^\infty f(\ep) \ep d\ep = \langle \ep (t) \rangle
\end{equation}
and
\begin{equation}
 \langle \ep^2 (t) \rangle = 3 \langle \ep (t) \rangle^2.
\end{equation}
Consequently, the variance and the standard deviation of the potential energy $\ep$ are
\begin{equation}
 \sigma^2(\ep (t)) = \langle \ep^2 (t) \rangle - \langle \ep (t) \rangle^2=2 \langle \ep (t) \rangle^2.
\end{equation}
and
\begin{equation}
 \sigma(\ep (t))=\sqrt{2} \langle \ep (t)\rangle
 \label{eq:ep-sigma}
\end{equation}
respectively.

Analogously, for the kinetic energy we have
\begin{eqnarray}
 f(\ek) & = &  \frac{1}{\sqrt{2\pi \langle \ek (t) \rangle}} \exp\left[ -\frac{\ek }{2\langle \ek (t) \rangle} \right] \times \frac{1}{\sqrt{\ek }}
\end{eqnarray}
leading to
\begin{equation}
 \mathcal{F}(\ek)=\mathrm{erf}\left[  \sqrt{\frac{\ek}{2\langle \ek (t) \rangle}} \; \right],
 \label{eq:ek-ccdf}
\end{equation}
and
\begin{equation}
  \sigma(\ek (t))=\sqrt{2} \langle \ek (t) \rangle,
  \label{eq:ek-sigma}
\end{equation}
where $\langle \ek (t) \rangle$ is given by Eq.~(\ref{eq:ek_wn}).

The full energy $\mathcal{E}$ is distributed according to
\begin{equation}
 f(\mathcal{E}) = \int_0^\mathcal{E} f(\mathcal{E}-\ek,\ek) d\ek.
\end{equation}
In order to find $f(\mathcal{E})$ the joint density $f(\ep,\ek)$ is required, which can be obtained from the $f(x,v)$ density which is a 2D normal distribution. Therefore, one needs to know the correlation matrix for $(x,v)$, see \cite{risken1984}.
Elements of the correlation matrix can be deduced from Eqs.~(\ref{eq:ek_wn}) and (\ref{eq:ep_wn}). The formula for the remaining element $\langle x(t) v(t) \rangle$ reads
\begin{equation}
 \langle x(t) v(t) \rangle = h\frac{1-\cos(2\omega t)}{4 m^2 \omega^2}.
\end{equation}
For the sake of clarity, we do not provide the formula for $f(\mathcal{E})$.
Nevertheless, in Fig.~\ref{fig:wn-cdf} ($b$) the exact $f(\mathcal{E})$ density is depicted as a solid line, see also \cite{czopnik2003frictionless}.

Figure~\ref{fig:n2wn-energy} presents standard deviations for kinetic and potential energies for the harmonic potential well. Solid lines present formulas given by Eqs.~(\ref{eq:ek_wn}) -- (\ref{eq:e_wn}), (\ref{eq:ep-sigma}) and (\ref{eq:ek-sigma}) while points correspond to results of computer simulations.
Please note that full and empty symbols of each type, i.e. squares and triangles, are superimposed.
Therefore, as predicted, the numerically estimated $\sigma(\mathcal{E}_{\dots} (t))$ is equal to $\sqrt{2} \langle \mathcal{E}_{\dots}(t)\rangle$.

%
%
\subsection{Non-harmonic potentials ($n>1$) \label{sec:nonharmonic}}

\subsubsection*{Gaussian white noise\label{sec:GWN}}

%
%

\begin{figure}[!h]%
 \centering
\includegraphics[angle=0,width=0.9\columnwidth]{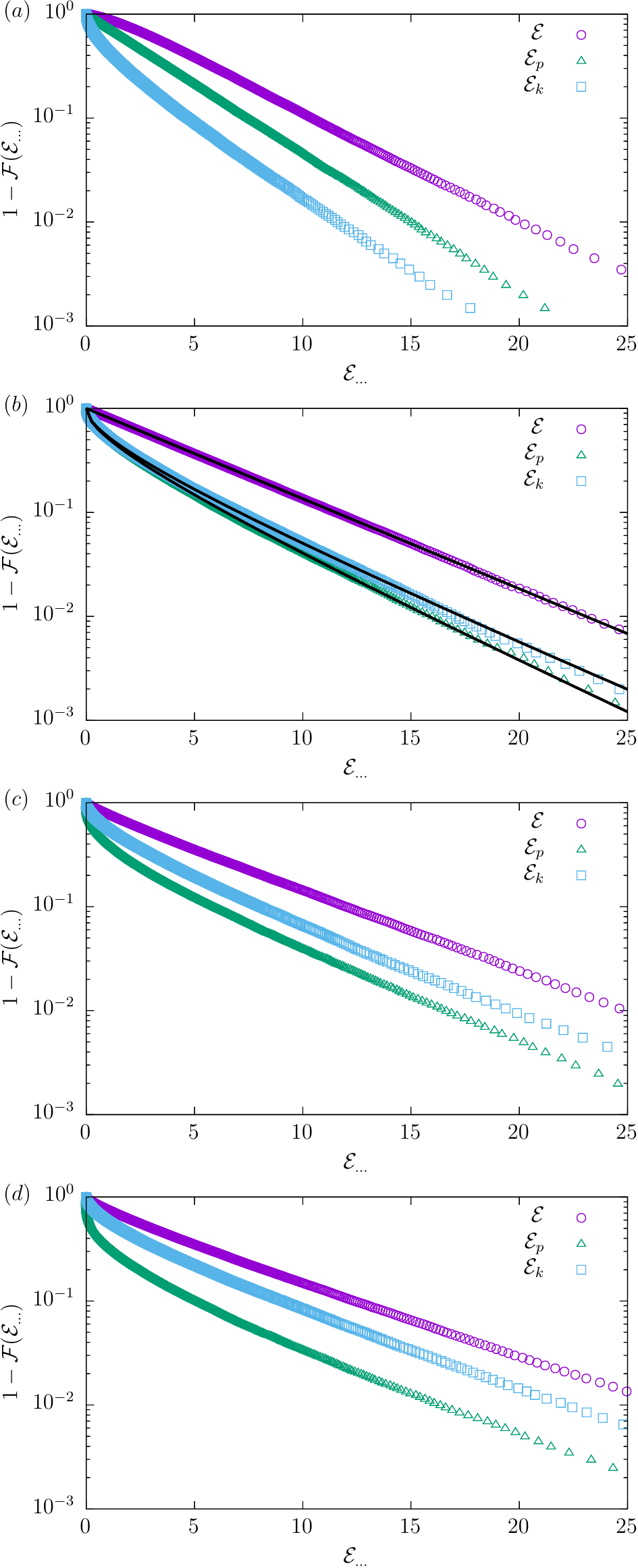} \\
 \caption{Energy distributions $f(\mathcal{E})$, $f(\ep)$ and $f(\ep)$ for different potentials: linear (top panel -- ($a$)), parabolic ($b$), cubic ($c$) and quartic (bottom panel -- ($d$)) with the GWN driving at a fixed $t=10$ time.
 Solid lines for the parabolic potential represent exact results, see Eqs.~(\ref{eq:ep-ccdf}), (\ref{eq:ek-ccdf}) and discussion in the text.}
 \label{fig:wn-cdf}
 \end{figure}

In the following section we focus our attention on the non-harmonic potentials i.e. $n>1$, see Eq.~(\ref{eq:potential}). We start with the Gaussian white noise driving. Next, we move to more general noises.
For the Gaussian white noise driving the results for the time dependence of the average total $\langle \mathcal{E}(t) \rangle$, potential $\langle \mathcal{E}_p(t) \rangle$ and kinetic $\langle \mathcal{E}_k(t) \rangle$ energies are presented in Fig.~\ref{fig:n2wn-all}.
Various rows (from top to bottom) correspond to various potential wells: linear, parabolic ($n=1$), cubic and quartic ($n=2$).
In Fig.~\ref{fig:n2wn-all}~($b$) analytical results, see Eqs.~(\ref{eq:ek_wn}) -- (\ref{eq:e_wn}) and \cite{mandrysz2018energetics} are compared with Monte Carlo (MC) simulations of the appropriate Langevin equation.
Analytical solutions
(\ref{eq:ek_wn}) -- (\ref{eq:e_wn})
have been constructed with the help of Eqs.~(\ref{eq:ep}) -- (\ref{eq:e}) and exact solution of Eq.~(\ref{eq:langevin}).
Numerical (Monte Carlo) results for undamped stochastic oscillators have been constructed by the algorithm presented in \cite{melbo2004numerical}. For the sake of simplicity, we have additionally assumed that $h=1$, $x(0)=0$ and $v(0)=0$.
As can be seen from Fig.~\ref{fig:n2wn-all} ($b$), for the parabolic potential, numerical simulations perfectly corroborate the theoretical predictions.

\begin{figure}[!h]%
	\centering
	  \includegraphics[angle=0,width=1.0\columnwidth]{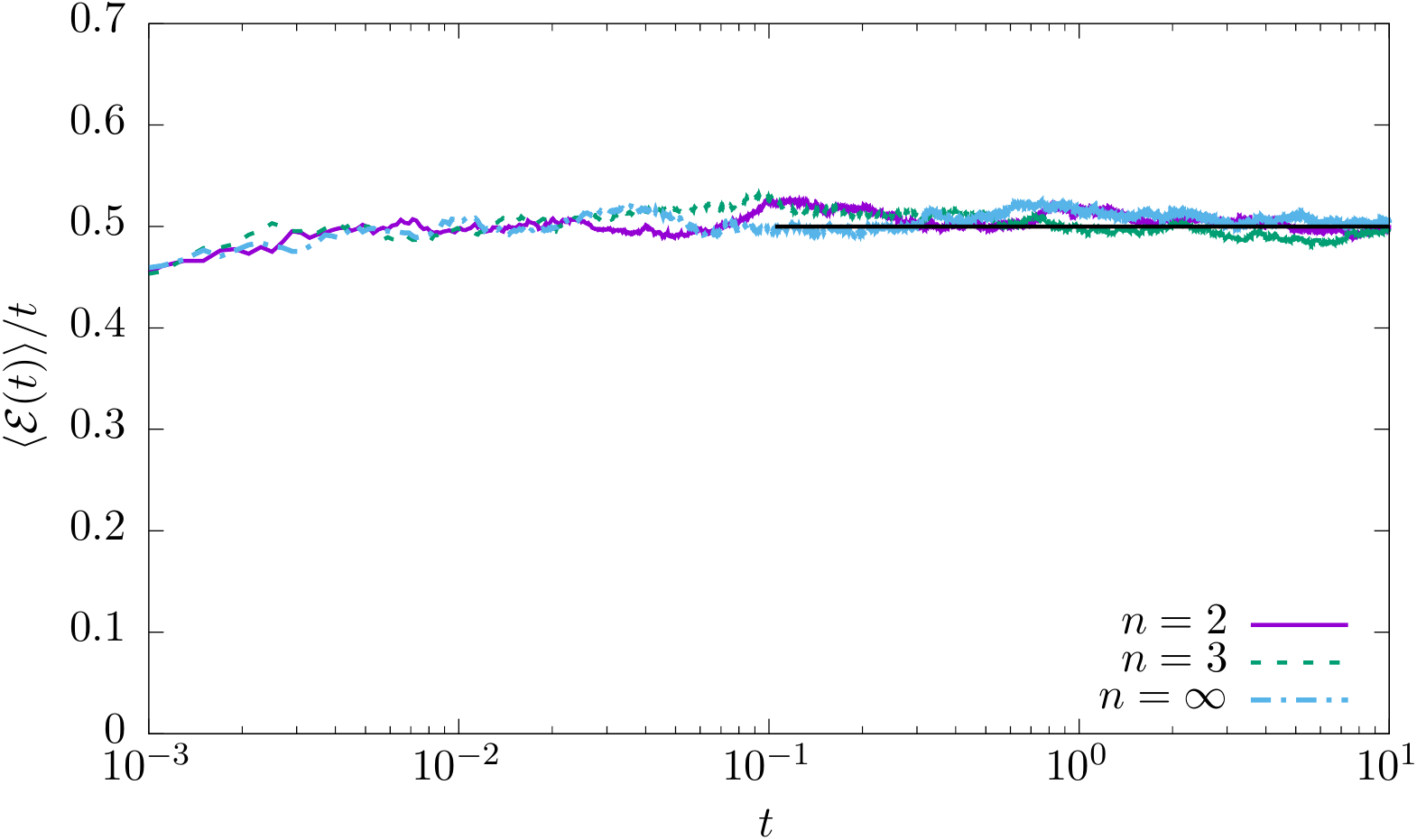} \\
	  \vspace{-0.5cm}
	\caption{Behavior of the rescaled (divided by $t$) average total energy $\langle \mathcal{E}(t) \rangle$ for different potential wells subject to the action of the GWN. The solid black line indicates the theoretical asymptotic prediction.}
	\label{fig:GWNgeneral}
\end{figure}

Surprisingly, comparing various plots in the left column of Fig.~\ref{fig:n2wn-all} one can see that for all considered potential wells with the Gaussian white noise driving the average total energy $\langle \mathcal{E}(t) \rangle$ exhibits the same time dependence. Using It\^o lemma \cite{mao2007stochastic} it is possible to confirm this observation in an analogous way to the damped harmonic oscillator \cite{yaghoubi2017energetics}.
Using the definition of the potential energy $\mathcal{E}_p=\frac{k}{2}x^2$ one gets
\begin{equation}
 \frac{d \mathcal{E}_p(x(t))}{dt}=kx(t) \frac{dx(t)}{dt}= k x(t) v(t).
\end{equation}
Therefore, after ensemble averaging the following formula is obtained
\begin{equation}
 \frac{d }{dt} \langle \mathcal{E}_p(t) \rangle= k \langle  x(t) v(t) \rangle,
 \label{eq:ep2}
\end{equation}
which is exactly the same as Eq.~(\ref{eq:ep}).
The kinetic energy $\mathcal{E}_k=\frac{1}{2}m v^2$ requires different treatment \cite{yaghoubi2017energetics}, because the velocity $v(t)$ fulfills the stochastic differential equation~(\ref{eq:set}).
Therefore, it is necessary to use the It\^o lemma
\begin{equation}
	\begin{aligned}
 d \mathcal{E}_k(v(t)) & = \frac{d \mathcal{E}_k}{dv} dv + \frac{1}{2}\frac{d^2 \mathcal{E}_k}{dv^2} (dv)^2+\dots \\
  & = m v dv + \frac{1}{2}m (dv)^2+\dots
 \label{eq:itolemma}
	\end{aligned}
\end{equation}
From Eq.~(\ref{eq:set})
\begin{eqnarray}
dv & = & -\omega^2 xdt+\sqrt{h}\xi(t)dt \\ \nonumber
& = & -\omega^2 xdt+\sqrt{h}dW(t),
\end{eqnarray}
where $dW(t)$ is the increment of the Wiener process.
Keeping terms that are at most linear in $dt$, $(dW(t))^2 = dt$, one gets
\begin{equation}
  \frac{d}{dt} \langle \mathcal{E}_k(t) \rangle = -k \langle x(t) v(t) \rangle + \frac{mh}{2}.
  \label{eq:ek2}
\end{equation}
After the addition of Eqs.~(\ref{eq:ep2}) and~(\ref{eq:ek2}) one obtains
\begin{equation}
  \frac{d}{dt} \langle \mathcal{E}(t) \rangle = \frac{mh}{2}.
  \label{eq:etot-der}
\end{equation}
Integration of Eq.~(\ref{eq:etot-der}) results in
\begin{equation}
   \langle \mathcal{E}(t) \rangle = \frac{mh}{2} \times t + \mathcal{E}_0,
  \label{eq:etot}
\end{equation}
where $\mathcal{E}_0$ is determined by the initial condition.
In an analogous way, it is possible to show that Eq.~(\ref{eq:etot}) holds for any single-well potential of $V(x)=k|x|^\nu/\nu$ ($\nu>0$) type when $\mathcal{E}_p=k|x|^\nu/\nu$.
In such a case $\langle x(t) v(t) \rangle$ in Eqs.~(\ref{eq:ep2}) and~(\ref{eq:ek2}) is to be replaced with $\langle x^{\nu-1}(t) v(t) \rangle$.
In a similar way, as for the parabolic potential terms $k\langle  x^{\nu-1}(t) v(t) \rangle$ cancel after the addition of Eqs.~(\ref{eq:ep2}) and~(\ref{eq:ek2}).
Therefore, inflow (pumping of energy), due to the contact with the thermal bath (described by the Gaussian white noise) results in the same (linear) time dependence of the average total energy, see Eq.~(\ref{eq:etot}).
This effect is very well visible in the left column of Fig.~\ref{fig:n2wn-all} and consequently in Fig.~\ref{fig:GWNgeneral} where the prediction given by Eq.~(\ref{eq:etot}) is further tested for $n \in \{2,3,\infty\}$ in the long-time limit.
Various rows in Fig.~\ref{fig:n2wn-all} present results for potentials with different values of the exponent~$n$: linear, parabolic ($n=1$), cubic and quartic ($n=2$), see Eq.~(\ref{eq:potential}). In all panels $\langle \mathcal{E}(t) \rangle$ is the same. Differences between all setups are recorded in the average potential $\langle \mathcal{E}_p(t) \rangle$ and average kinetic $\langle \mathcal{E}_k(t) \rangle$ energies which display very different time dependence.
Differences between various types of single-well potentials are also visible in the characteristics of energy distributions: median (quantile $q_{0.5}(t)$) and width (defined as inter-quantile width -- $q_{0.9}(t)-q_{0.1}(t)$), which are presented in the middle and right columns of Fig.~\ref{fig:n2wn-all}.

Further differences between various potentials are inspected in Fig.~\ref{fig:wn-cdf} which presents energy distributions $f(\mathcal{E})$, $f(\ep)$ and $f(\ek)$ at $t=10$ for the potential wells studied in Fig.~\ref{fig:n2wn-all}.
For the parabolic potential ($n=2$) solid lines represent exact results which perfectly agree with results of Monte Carlo simulations, see Fig.~\ref{fig:wn-cdf} ($b$).
Since characteristics of energy distributions depicted in Fig.~\ref{fig:n2wn-all} differ, likewise energy distributions depicted in Fig.~\ref{fig:wn-cdf} depend on the potential type, see the next subsection.

The linear growth of the average total energy $\langle \mathcal{E}(t) \rangle$ for any single-well potential perturbed by the GWN raises a question whether the observed effect, i.e. the linear growth of $\langle \mathcal{E}(t) \rangle$, holds for other types of noise.
In general, the answer is negative and depends on the steepness of the potential.
Nevertheless, as we show later, in the long-time limit it is possible to fine-tune the rate of energy growth by adjusting noise parameters, as is done in the following case of OUN.

The Gaussian white noise can be generalized to the $\alpha$-stable (L\'evy type) white noise, see \cite{janicki1994,dubkov2008}.
For a non-equilibrium noise of $\alpha$-stable type the general property visible for the GWN is no longer true.
Moreover, for the harmonic potential well time dependent densities $f(x,v)$ are given by 2D $\alpha$-stable densities \cite{samorodnitsky1994,sokolov2010} which are characterized by the diverging variance, and possibly also by the diverging mean.
Therefore, it is necessary to use different measures, e.g. robust measures based on quantiles of the energy distribution.
These measures, by analogy with medians and inter-quantile widths presented in Fig.~\ref{fig:n2wn-all}, confirm the dependence of energetic properties of stochastic oscillators driven by an $\alpha$-stable noise both on the noise type and the potential type.

\subsubsection*{Ornstein-Uhlenbeck and Markovian dichotomous noises\label{sec:OU}}

\begin{figure}[!h]%
\centering
\includegraphics[angle=0,width=1.0\columnwidth]{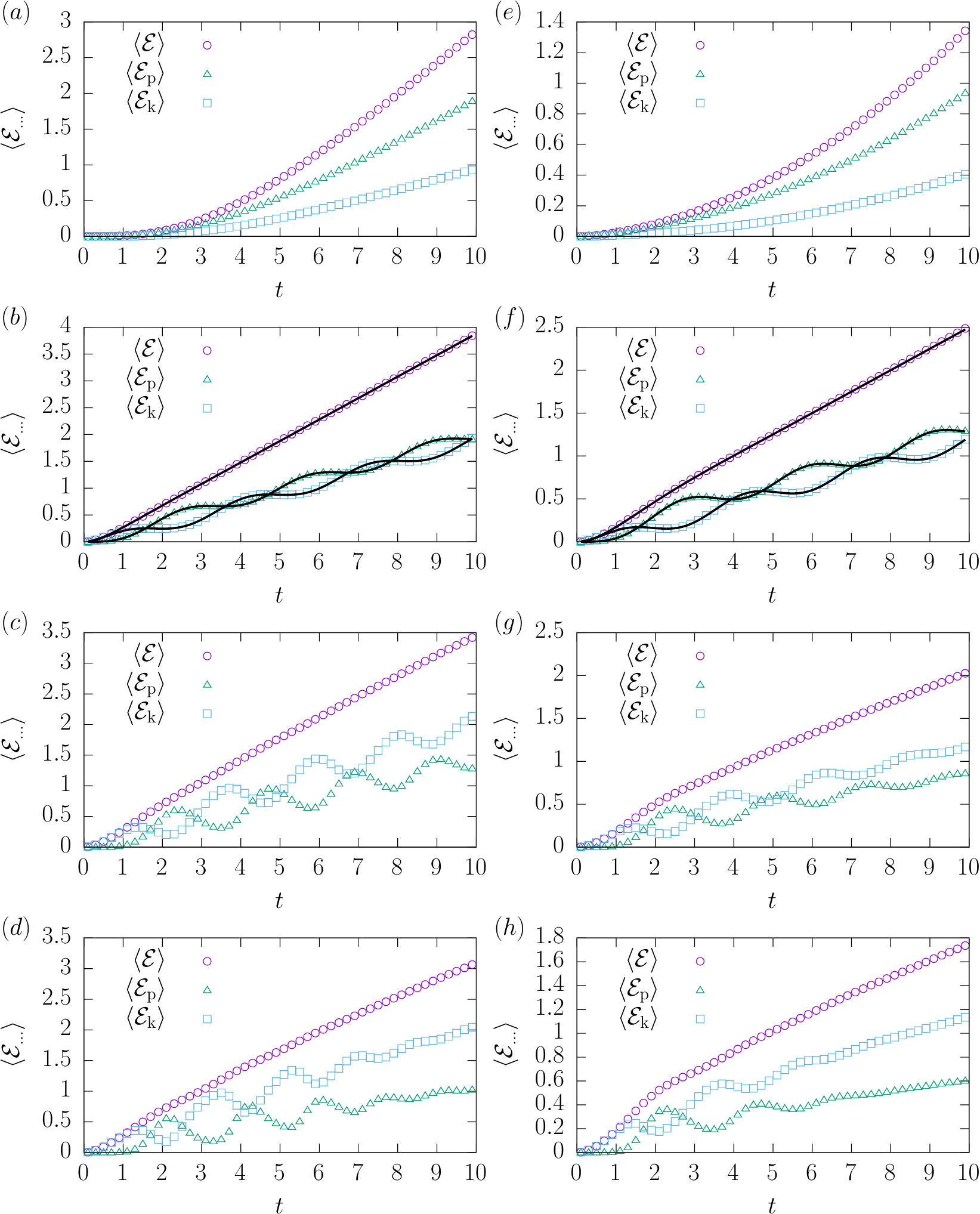} \\
\caption{Average energies $\langle \mathcal{E}_{\dots} (t) \rangle$ for DN (left column -- ($a$) -- ($d$)) and OUN (right column -- ($e$)--($h$)) drivings for linear (top panel -- ($a$) and ($e$)), parabolic, cubic and quartic (bottom panel -- ($d$) and ($h$)) potentials.
Solid lines for the parabolic potential ($b$) and ($f$) depict exact results, see \cite{mandrysz2018energetics}.
}
\label{fig:dn-ou}
\end{figure}

\begin{figure}[!h]%
\centering
\includegraphics[angle=0,width=0.9\columnwidth]{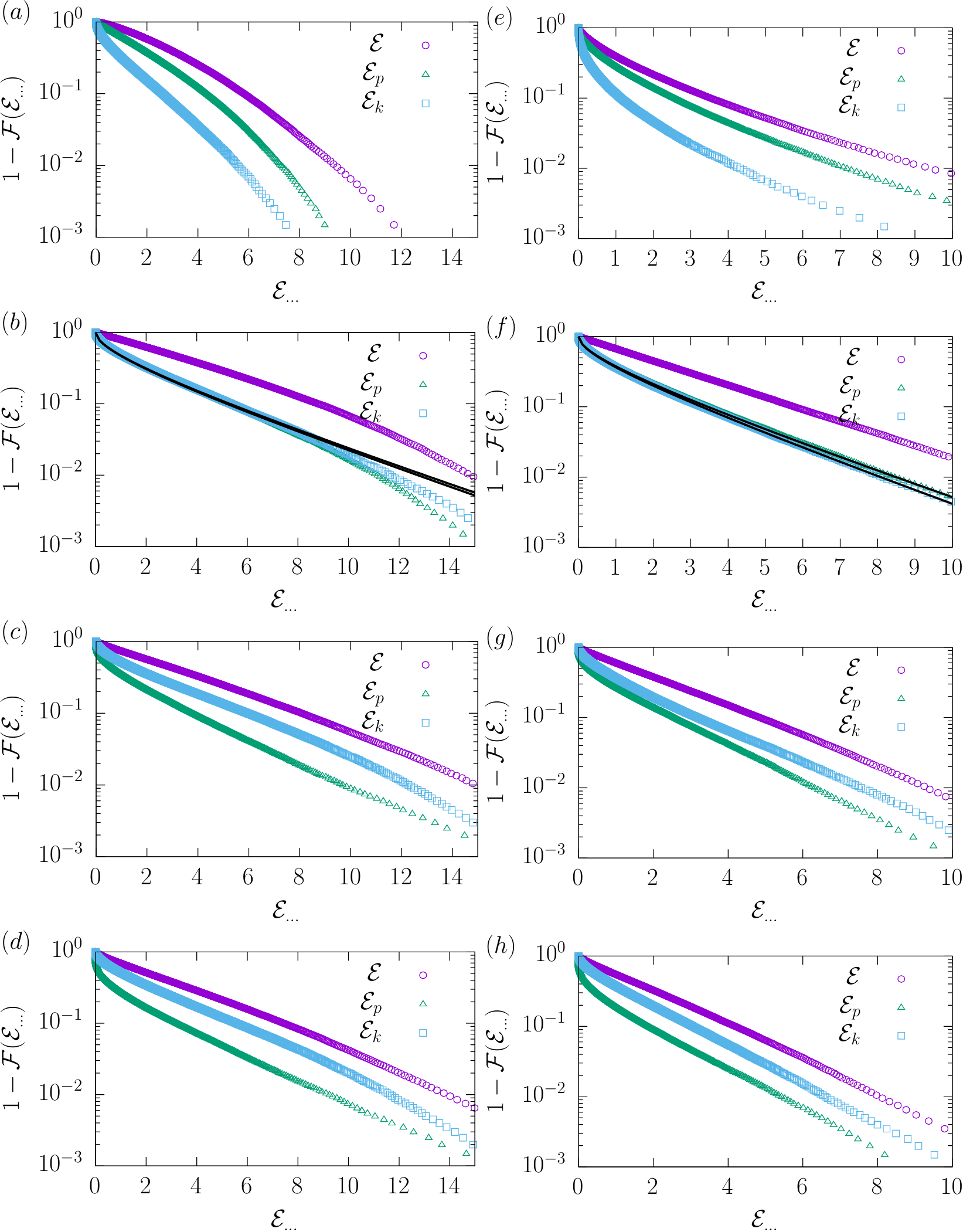} \\
 \caption{Energy distributions for DN (left column -- ($a$)--($d$)) and OUN (right column -- ($e$)--($h$)) drivings for linear (top panel -- ($a$) and ($e$)), parabolic, cubic and quartic (bottom panel -- ($d$) and ($h$)) potentials at a fixed $t=10$ time.
 }
 \label{fig:dn-oun-cdf}
 \end{figure}

We now proceed to study energetic properties of stochastic oscillators driven by colored noises, e.g. Ornstein-Uhlenbeck and Markovian dichotomous noises.
Fig.~\ref{fig:dn-ou} presents $\langle \mathcal{E}(t) \rangle$ for linear, parabolic ($n=2$), cubic and quartic ($n=2$) potential wells. The left column corresponds to the DN driving while the right one corresponds to the OUN driving.
Fig.~\ref{fig:dn-ou} clearly shows that time dependence of the average total energy $\langle \mathcal{E}(t) \rangle$ depends both on the noise and potential types.
This is a consequence of a lack of whiteness in the driving noise.
Additional differences between various considered setups are depicted in Fig.~\ref{fig:dn-oun-cdf} which shows complementary cumulative densities of energy.
Fig.~\ref{fig:dn-oun-cdf} presents energy distributions for various potential types.
Solid lines in the second row, Fig.~\ref{fig:dn-oun-cdf} ($b$) and ($f$), represent exact results for the parabolic potential under the Gaussian white noise with the formulas for $\langle \ep (t) \rangle$, $\langle \ek (t) \rangle$ and $\langle \mathcal{E} (t) \rangle $ for appropriate noises, see Eqs.~(\ref{eq:e_dn_as}) -- (\ref{eq:e_oun_as}).
Consequently, solid lines demonstrate how the harmonic oscillator driven by a colored noise (points) differs from its Gaussian counterpart (solid lines).

\begin{figure}[!htbp]%
	\centering
	  \includegraphics[angle=0,width=1.0\columnwidth]{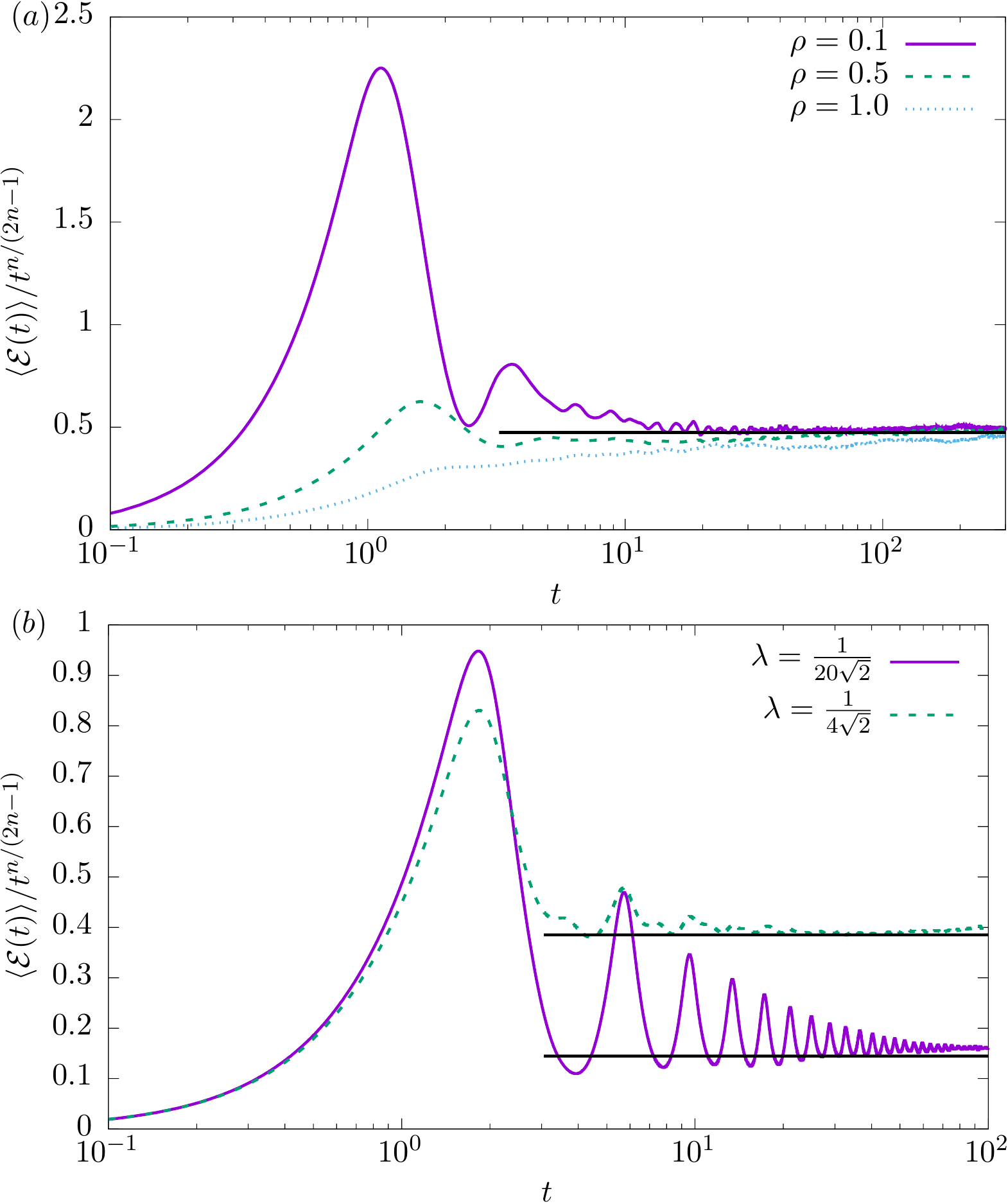} \\
	  \vspace{-0.5cm}
	\caption{Behavior of total rescaled energy for OUN (top panel -- ($a$)) and DN (bottom panel -- ($b$)) drivings for $n=3$. Solid black lines indicate theoretical asymptotic predictions.}
\label{fig:generalN}
\end{figure}

The problem of the general stochastic oscillator driven by OUN was studied in \cite{mallick2005anharmonic} where the formula for the evolution of the average mechanical energy $\langle \mathcal{E} (t) \rangle$ was derived
\begin{equation}
	\langle \mathcal{E}_{OUN}(t) \rangle = \frac{\Gamma{\left(\frac{3 n + 1}{4 n - 2}\right)}}{\Gamma{\left(\frac{n + 1}{4 n - 2}\right)}} \left[ \frac{(2 n - 1)^2}{2 n^2} \mu t \right]^{n/(2n-1)},
	\label{eq:oun-general}
\end{equation}
where
\begin{equation}
	\mu = (2 n)^{1/n} \frac{\Gamma{\left(\frac{3}{2 n}\right)} \Gamma{\left(\frac{n + 1}{2 n}\right)}}{\Gamma{\left(\frac{1}{2 n}\right)} \Gamma{\left(\frac{n + 3}{2 n}\right)}}.
\end{equation}
Eq.~(\ref{eq:oun-general}) clearly indicates that the exponent characterizing the growth of the average total energy depends on the steepness of the potential $n$.
At this point we would also like to note that for $\mathcal{D}=1$, see Eq.~(\ref{eq:ou-ac}), the resulting energy growth rate, i.e. the prefactor in Eq.~(\ref{eq:oun-general}), does not depend on the damping rate $\rho$ (inverse of the correlation time).
Additionally, in \cite{mallick2003scaling} the following relations have been obtained:
\begin{equation}
	\langle \mathcal{E}(t) \rangle  = \frac{n+1}{2n} \langle \dot{x}^2 (t) \rangle, \\
\end{equation}
and
\begin{equation}
	\langle \dot{x}^2 (t) \rangle  = \langle x^{2n} (t) \rangle
\end{equation}
which provide the relation between the growth of average kinetic and potential energies.
The solution for the Ornstein-Uhlenbeck driving was reported \cite{mallick2003scaling} 
to hold also for the symmetric Markovian dichotomous noise, which in our case yields:
\begin{equation}
	\langle \mathcal{E}_{DN}(t) \rangle = \frac{\Gamma{\left(\frac{3 n + 1}{4 n - 2}\right)}}{\Gamma{\left(\frac{n + 1}{4 n - 2}\right)}} \left[ \frac{(2 n - 1)^2}{2 n^2} 4 \lambda \mu t \right]^{n/(2n-1)}.
	\label{eq:dn-general}
\end{equation}
At this point, we would like to underline that for an appropriate choice of parameters, despite a different character of Markovian dichotomous and Ornstein-Uhlenbeck noises, both noises could result in the same asymptotic scaling of average total energies, i.e. $\langle \mathcal{E}_{OUN}(t) \rangle/t^{n/(2n-1)}$ tends to the same limit as $ \langle \mathcal{E}_{DN}(t) \rangle/t^{n/(2n-1)}$.
More precisely, in order to reach the same scaling, it is necessary to choose such parameters that autocorrelation functions of both noises, which are given by Eqs.~(\ref{eq:dn-ac}) and Eq.~(\ref{eq:ou-ac}), are the same.
For $n=3$, predictions given by Eqs.~(\ref{eq:oun-general}) and (\ref{eq:dn-general}) have been tested using Monte Carlo simulations, see Fig.~\ref{fig:generalN}.
MC tests have proven that at sufficiently large time $t$ scaling predicted by Eqs.~(\ref{eq:oun-general}) and (\ref{eq:dn-general}) is reached.

%
%
\subsection{Infinite rectangular potential well ($n=\infty$)}

\begin{figure}[!htbp]%
	\centering
	  \includegraphics[angle=0,width=1.0\columnwidth]{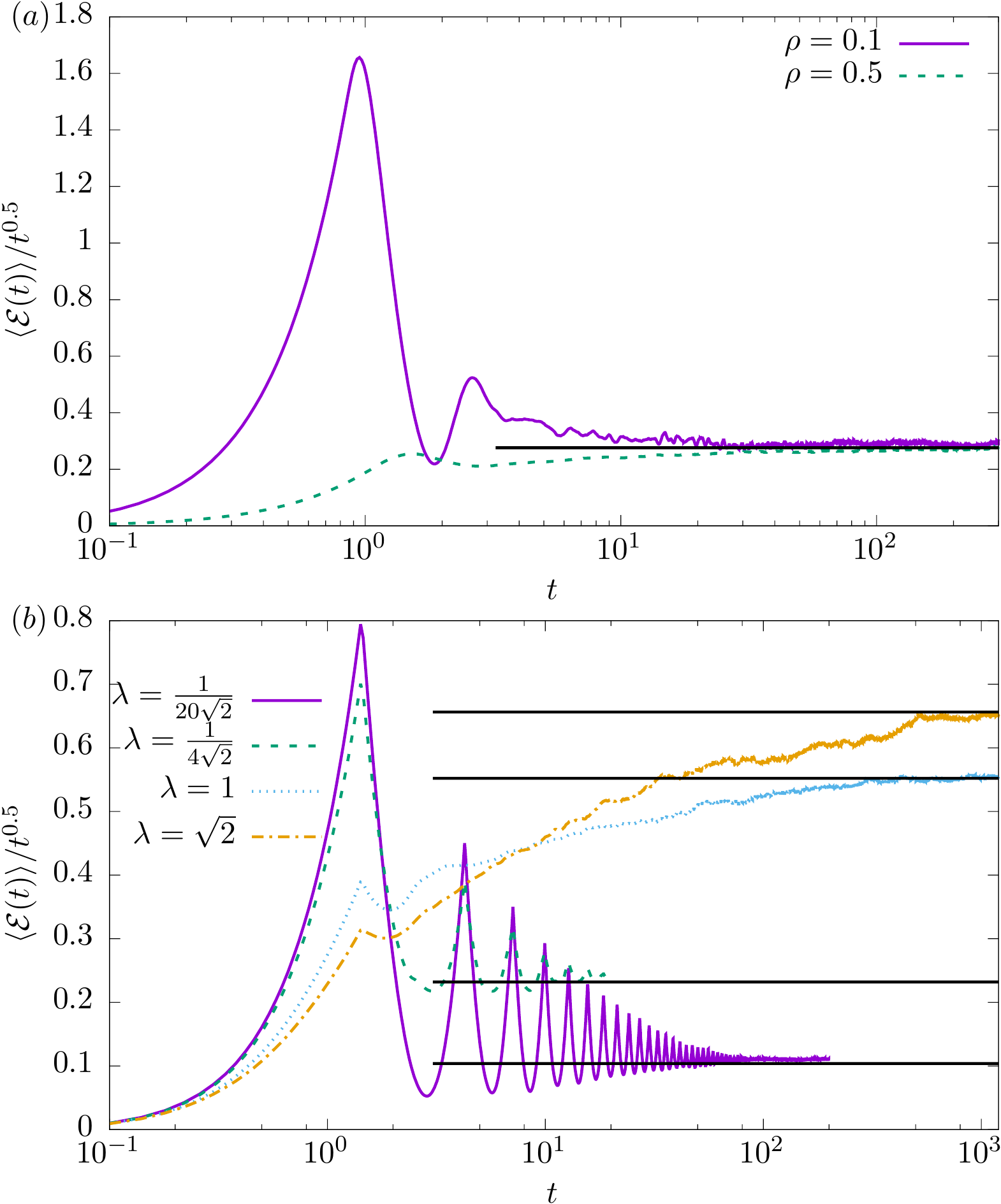} \\
	  \vspace{-0.5cm}
	\caption{Behavior of total rescaled energy for OUN (top panel -- ($a$)) and DN (bottom panel -- ($b$)) in the infinite rectangular potential well ($n=\infty$). Solid black lines indicate theoretical asymptotic predictions.}
\label{fig:generalINF}
\end{figure}

In the limit of $n\to\infty$ the potential well of $x^{2n}/2n$ type transforms into the infinite rectangular potential well. A particle moving in the infinite rectangular potential well, except time instants when it reflects from the boundary,
moves like a free particle. As we already noted, for the GWN the average energy scales
linearly in time for any single-well potential,
see Fig.~\ref{fig:GWNgeneral} and Eq.~(\ref{eq:etot}).
The time dependence of the rescaled energy for $n=\infty$ for systems driven by Markovian dichotomous noise and Ornstein-Uhlenbeck noise are presented in Fig.~\ref{fig:generalINF}.
From Fig.~\ref{fig:generalINF} it can be deducted that in the $n\to\infty$ limit predictions given by Eqs.~(\ref{eq:oun-general}) and (\ref{eq:dn-general}), see \cite{mallick2003scaling}, are valid.

The presence of boundaries affects the scaling predicted by Eq.~(\ref{eq:free-dn}).
For example, the Markovian symmetric dichotomous noise stays constant for exponentially distributed time $\tau$, i.e.
\begin{equation}
 p(\tau) = \lambda \exp(-\lambda \tau)
\end{equation}
with average time $\langle \tau \rangle$ given by
\begin{equation}
 \langle \tau \rangle = \frac{1}{\lambda}.
\end{equation}
At $\lambda \to 0$ the Markovian dichotomous process stays constant.
Therefore, the motion of a particle is like a free fall of a bouncing ball.
It moves in the direction of the randomly chosen boundary (floor in case of the bouncing ball), i.e. to $x=\pm 1$ where the boundaries are located.
After reflection at the boundary, the velocity is reversed and the particle returns to the origin that is to its starting point.
At the origin the motion is stopped and reversed by the external force.
In the chosen setup, the time needed to reflect for the first time is $\sqrt{2}$. The particle returns to its starting point after double the time, i.e. $2\sqrt{2}$.
In such a case ($\lambda=0$) the motion is fully periodic with the period $T=2\sqrt{2}$. The particle interacts with the one boundary only which is selected by the initial value of the dichotomous noise.
More precisely, for $\xi_{DN}(0)=+1$ the particle reflects from the right boundary ($x=+1$) only, while for $\xi_{DN}(0)=-1$ from the left boundary ($x=-1$) only.

\begin{figure}[!h]%
\centering
\begin{tabular}{c}
\includegraphics[angle=0,width=1.0\columnwidth]{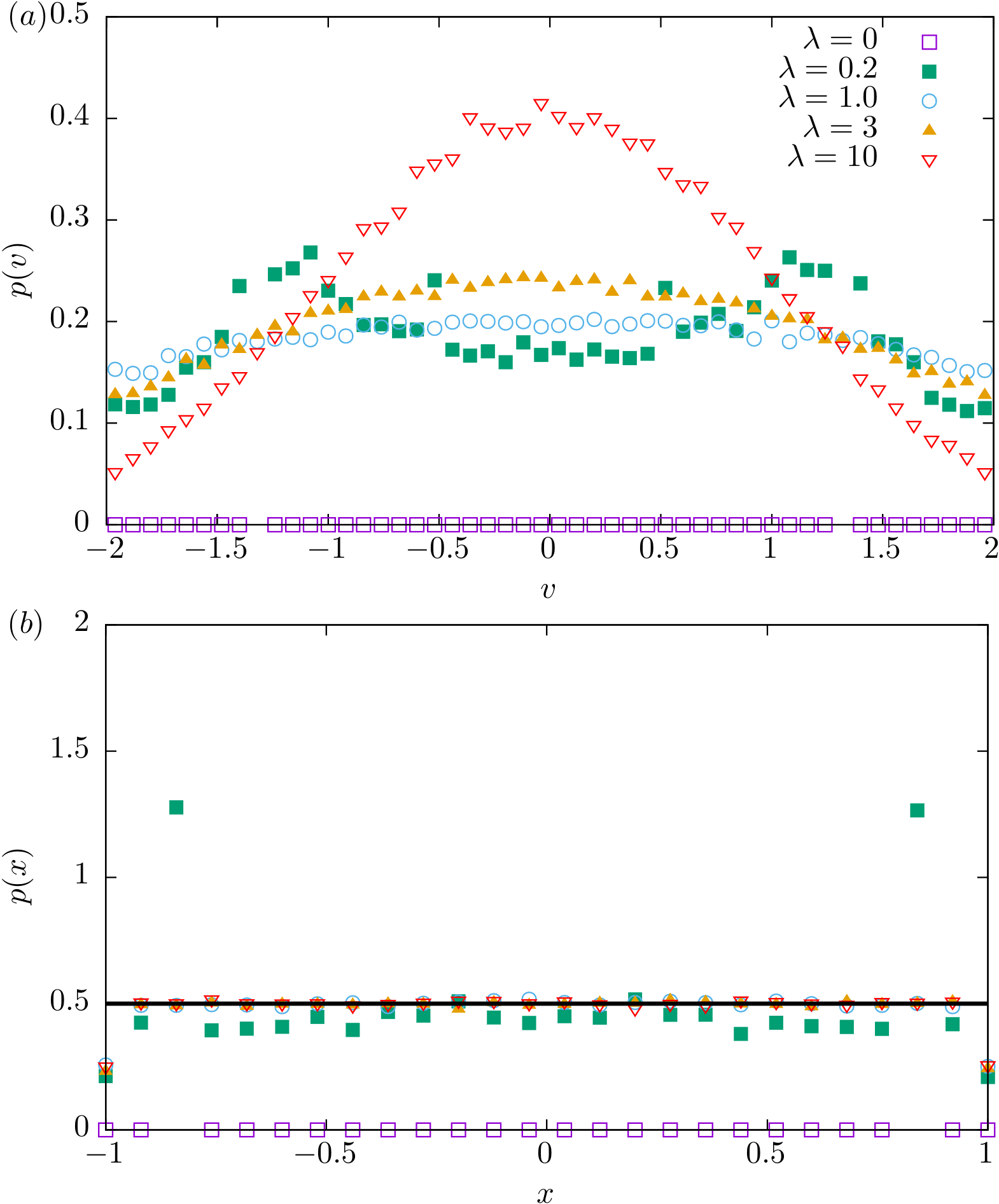} \\
\end{tabular}
\caption{Velocity $p(v)$ (top panel -- ($a$)) and position $p(x)$ (bottom panel -- ($b$)) histograms at a fixed time $t=10$ for the infinite rectangular potential well and dichotomous noise with various switching rates~$\lambda$.
For the legend see panel ($a$).
}
\label{fig:rw-xv-his}
\end{figure}

\begin{figure}[!h]%
\centering
\begin{tabular}{c}
\includegraphics[angle=0,width=1.0\columnwidth]{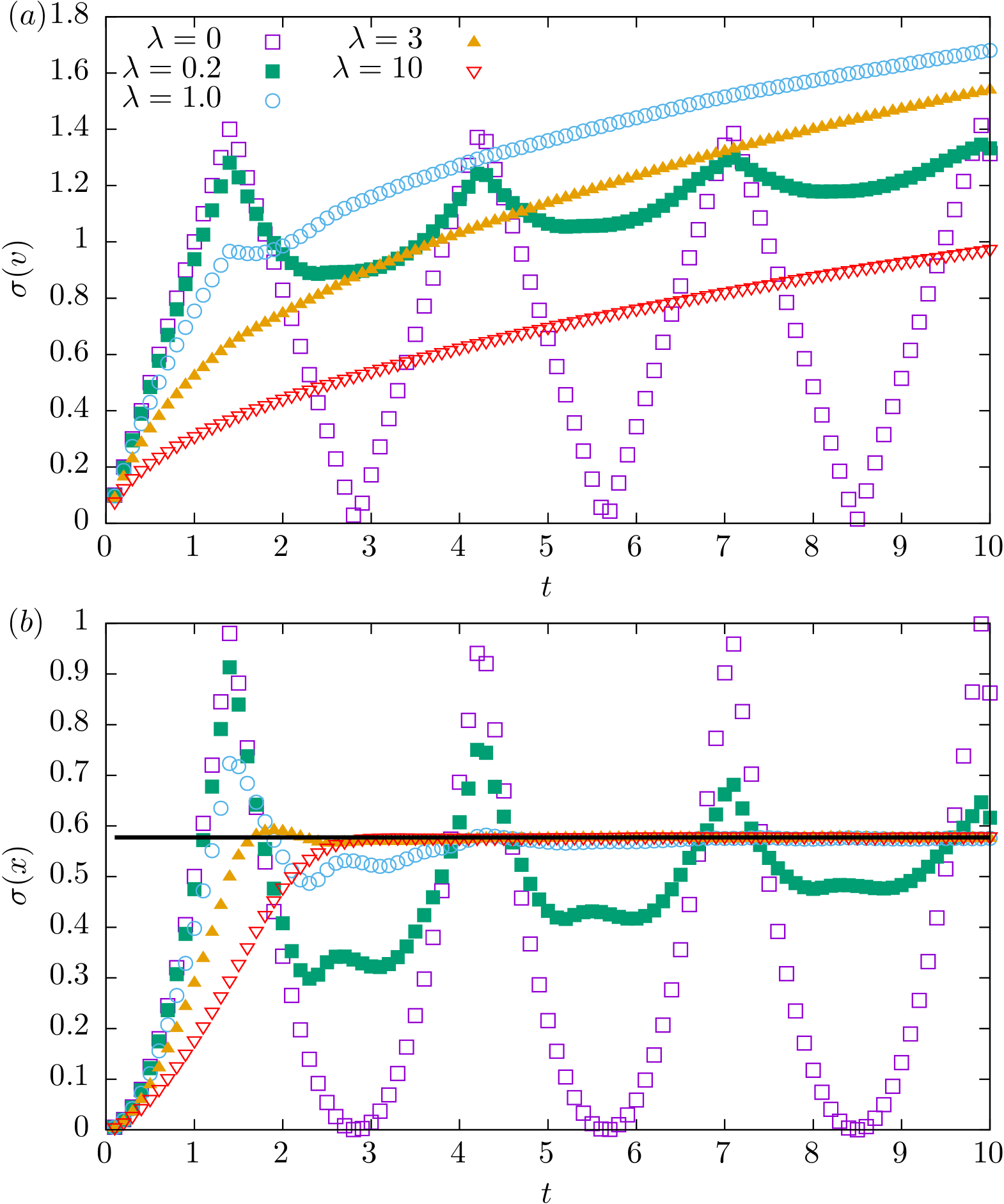} \\
\end{tabular}
\caption{Standard deviation of velocity $\sigma(v)$ (top panel -- ($a$)) and position $\sigma(x)$ (bottom panel -- ($b$)) for the infinite rectangular potential well and the Markovian dichotomous noise with various switching rates~$\lambda$.
For the legend see panel ($a$).
}
\label{fig:rw-xv-stdev}
\end{figure}

For $\lambda=0$ the motion is fully deterministic thus the position $p(x)$ and velocity $p(v)$ densities consist of moving delta peaks at the deterministic velocity $v(t)$ and the deterministic position $x(t)$, see Fig.~\ref{fig:rw-xv-his}. Due to the initial condition set at the DN, i.e. $\xi_{DN} = \pm 1$, for $\lambda=0$, there are two symmetric peaks in $p(v)$ and $p(x)$ densities.
With the increasing switching rate $\lambda$ the particle starts to change its direction due to noise and the peaks smear out.
For a large switching rate $\lambda$, the velocity distribution resembles normal density while
the position distribution becomes uniform on $[-1,1]$, see Fig.~\ref{fig:rw-xv-his}~($b$).
Therefore, $\sigma(x)$ tends to $1/\sqrt{3} \approx 0.58$, see Fig.~\ref{fig:rw-xv-stdev}~($b$).

Boundaries ``modulate'' how the energy is pumped into the system over short times. By adjusting the correlation time of the DN one can control the rate of the delivered energy at long times, see Fig.~\ref{fig:generalINF}~($b$).
The increase in the switching rate $\lambda$ destroys periodicity of $\langle \ek (t) \rangle$. For a large enough $\lambda$ the average kinetic energy grows in time, but the average level of energy reached at a fixed time is a monotonous function of the switching rate $\lambda$ only in the long-time limit, see Fig.~\ref{fig:generalINF}~($b$).
In contrast, for a finite time $t$ it can be non-monotonous as confirmed by crossing lines in Fig.~\ref{fig:generalINF}~($b$).
As can be seen in Fig.~\ref{fig:generalINF}~($a$), one finds perfect agreement with the scaling predicted by Eq.~(\ref{eq:oun-general}).
For a finite $\lambda$, the average energy $\langle \mathcal{E}(t) \rangle$ scales asymptotically like $t^{1/2}$.
Nevertheless, another special limit should be discussed. The symmetric Markovian dichotomous noise reduces to the Gaussian white noise in the limit of $\lambda \to \infty$ under the additional condition that noise values, here set to $\pm 1$, also tend to infinity, see \cite{broeck1983,bena2006}.
Consequently, for a sufficiently large $\lambda$ the average energy scales in the same manner as for the GWN, i.e. $\langle \mathcal{E}(t) \rangle \propto t$, however the proportionality coefficient depends on $\lambda$, i.e. it is $1/2\lambda$, because values of the Markovian dichotomous process are kept constant.
Despite the presence of the reflecting boundaries, this scaling is the same as for a free particle.
The transition from $t^{1/2}$ to $t$ scaling of the average energy is due to the vanishing correlation time.

The energy of a particle trapped in the infinite rectangular potential well subject to the action of DN grows slower when the noise changes its state less often, i.e. the faster the switching rate,
the higher the coefficient (prefactor) with which the energy grows in the $t \to \infty$ limit.
At the same time the rescaled energy saturates slower, i.e. a longer time is necessary to reach the asymptotic dependence.
Thus, in principle, one could control the amount of energy pumped over short times by modifying the width of the infinite rectangular potential well.
At long times the amount of delivered energy can be adjusted by the correlation time.

\section{Summary and Conclusions\label{sec:summary}}

We have studied the undamped motion in single-well potentials of $k x^{2n}/2n$ type subject to the action of stochastic driving.
The absence of the damping term breaks the energy balance because there is no dissipation in the system.
Due to the presence of noise, energy is pumped into the system.
Therefore, both average kinetic and potential energies grow in time.
The examination of the energy growth curve in single-well potential wells constituted the main subject of this research.

First of all, we have shown that undamped motion perturbed by the Gaussian white noise, in any single-well potential, results in the linear scaling of the average total energy.
At the same time, the dependence of average kinetic and potential energies is sensitive to the potential type.

Furthermore, we have considered other, non-white, noise types, i.e. the symmetric Markovian dichotomous noise and the Ornstein-Uhlenbeck noise.
For these two special types of colored noises, average energies scale in time with the potential dependent exponent different from that of the Gaussian white noise.

The limiting case of the infinite potential steepness ($n \to \infty$) has also been studied.
In such a case, the potential of $kx^{2n}/2n$ type reduces to the infinite rectangular potential well.
Therefore, the motion of the particle is affected by external forces only during collision events with the ideally reflecting boundaries.
These collisions result in hard velocity reversals, i.e. the velocity, which typically differs from zero, changes its sign at the boundary.
For the increasing $n$, the predicted scaling of the average energy recorded for a finite $n$ stays valid under the condition that the noise correlation time stays finite.
In the special limit of a vanishing correlation time the Markovian symmetric dichotomous noise and the Ornstein-Uhlenbeck noise can be reduced to the Gaussian white noise.
Therefore, for a very small correlation time the scaling of the average energy predicted for the Gaussian noise is recovered.
The dependence of the average energy scaling on details of the system dynamics opens potential practical applications.
In particular, the undamped motion in single-well potentials can be used to identify the underlying noise type.
Finally, the reintroduction of the damping term results in a situation when the dissipation of energy prevents average energies from an unlimited growth and the departure from the studied regime.

\begin{acknowledgments}
 This project was supported by the National Science Center grant (2014/13/B/ST2/02014).
Computer simulations have been performed at the Academic
Computer Center Cyfronet, AGH University of Science and Technology (Krak\'ow, Poland)
under CPU grant ``DynStoch''.

\end{acknowledgments}


\begin{thebibliography}{10}

\bibitem{sekimoto2010stochastic}
K. Sekimoto, {\em Stochastic Energetics} ({Springer Verlag}, Berlin, 2010),
  Vol.~799.

\bibitem{seifert2012stochastic}
U. Seifert, Rep. Prog. Phys. {\bf 75},  126001  (2012).

\bibitem{gammaitoni1998}
L. Gammaitoni, P. H\"anggi, P. Jung, and F. Marchesoni, Rev. Mod. Phys. {\bf
  70},  223  (1998).

\bibitem{doering1992}
C.~R. Doering and J.~C. Gadoua, Phys. Rev. Lett. {\bf 69},  2318  (1992).

\bibitem{reimann2002}
P. Reimann, Phys. Rep. {\bf 361},  57  (2002).

\bibitem{joubaud2007fluctuation}
S. Joubaud, N. Garnier, and S. Ciliberto, J. Stat. Mech. {\bf 2007},  P09018
  (2007).

\bibitem{hwalisz1989colored}
L. H'walisz, P. Jung, P. H\"anggi, P. Talkner, and L. {Schimansky-Geier}, Z.
  Phys. B {\bf 77},  471  (1989).

\bibitem{metzler2000}
R. Metzler and J. Klafter, Phys. Rep. {\bf 339},  1  (2000).

\bibitem{metzler2004}
R. Metzler and J. Klafter, J. Phys. A: Math. Gen. {\bf 37},  R161  (2004).

\bibitem{czopnik2003frictionless}
R. Czopnik and P. Garbaczewski, Physica A {\bf 317},  449  (2003).

\bibitem{bena2006}
I. Bena, Int. J. Mod. Phys. B {\bf 20},  2825  (2006).

\bibitem{eab2018ornstein}
C.~H. Eab and S. Lim, Physica A {\bf 492},  790  (2018).

\bibitem{reichl1998}
L.~E. Reichl, {\em A modern course in statistical physics} (John Wiley, New
  York, 1998).

\bibitem{reif2009}
F. Reif, {\em Fundamentals of statistical and thermal physics} (Waveland Press,
  Long Grove, 2009).

\bibitem{lin2011undamped}
N. Lin and S. Lototsky, Comm. Stoch. Anal. {\bf 5},  13  (2011).

\bibitem{mandrysz2018energetics}
M. Mandrysz and B. Dybiec, Acta Phys. Pol. B {\bf 49},  871  (2018).

\bibitem{chechkin2008introduction}
A.~V. Chechkin, R. Metzler, J. Klafter, and V.~Y. Gonchar,  in {\em Anomalous
  transport: Foundations and applications}, edited by R. Klages, G. Radons, and
  I.~M. Sokolov (Wiley-VCH, Weinheim, 2008), pp.\ 129--162.


\bibitem{mallick2005anharmonic}
K. Mallick and P. Marcq, J. Stat. Phys. {\bf 119},  1  (2005).

\bibitem{gitterman2005noisy}
M. Gitterman, {\em The Noisy Oscillator: The First Hundred Years, from
  {{Einstein}} until Now} (World Scientific Publishing, Singapore, 2005).

\bibitem{gitterman2013noisy}
M. Gitterman, {\em The Noisy Oscillator: Random Mass, Frequency, Damping}
  (World Scientific Publishing, Singapore, 2013).

\bibitem{kubo1966fluctuation}
R. Kubo, Rep. Prog. Phys. {\bf 29},  255  (1966).

\bibitem{risken1984}
H. Risken, {\em The {Fokker-Planck} equation. Methods of solution and
  application} (Springer Verlag, Berlin, 1984).

\bibitem{horsthemke1984}
W. Horsthemke and R. Lefever, {\em Noise-inducted transitions. Theory and
  applications in physics, chemistry, and biology} (Springer Verlag, Berlin,
  1984).

\bibitem{Note1}
This definition does converge to the GWN for $\protect \mathcal {D}=\rho $ and
$\rho \to \infty $.

\bibitem{gillespie1996exact}
D.~T. Gillespie, Phys. Rev. E {\bf 54},  2084  (1996).

\bibitem{mao2007stochastic}
X. Mao, {\em Stochastic Differential Equations and Applications} ({Woodhead
  Publishing}, Oxford, 2007).

\bibitem{tome2015stochastic}
T. Tom\'e and M.~J. {de Oliveira}, {\em Stochastic {{Dynamics}} and
  {{Irreversibility}}} ({Springer Verlag}, Berlin, 2015).

\bibitem{melbo2004numerical}
A.~H. Melb\o and D.~J. Higham, Appl. Numer. Math. {\bf 51},  89  (2004).

\bibitem{yaghoubi2017energetics}
M. Yaghoubi, M.~E. Foulaadvand, A. B\'erut, and J. \L{}uczka, J. Stat. Mech.
  {\bf 2017},  113206  (2017).

\bibitem{janicki1994}
A. Janicki and A. Weron, {\em Simulation and chaotic behavior of
  $\alpha$-stable stochastic processes} (Marcel Dekker, New York, 1994).

\bibitem{dubkov2008}
A.~A. Dubkov, B. Spagnolo, and V.~V. Uchaikin, Int. J. Bifurcation Chaos. Appl.
  Sci. Eng. {\bf 18},  2649  (2008).

\bibitem{samorodnitsky1994}
G. Samorodnitsky and M.~S. Taqqu, {\em Stable non-{Gaussian} random processes:
  Stochastic models with infinite variance} (Chapman and Hall, New York, 1994).

\bibitem{sokolov2010}
I.~M. Sokolov, B. Dybiec, and W. Ebeling, Phys. Rev. E {\bf 83},  041118
  (2011).

\bibitem{mallick2003scaling}
K. Mallick and P. Marcq, Eur. Phys. J. B {\bf 31},  553  (2003).

\bibitem{broeck1983}
C. Van~den Broeck, J. Stat. Phys. {\bf 31},  467  (1983).

\end{thebibliography}

\def\url#1{}

\end{document}